\documentclass[12pt]{article}
\usepackage{tikz}
\usepackage{amsmath}
\usepackage{mathtools}
\usepackage{fancyhdr}
\usepackage{gensymb}
\usepackage{graphicx}
\usepackage{cite}
\usetikzlibrary{decorations.pathmorphing, arrows, patterns}
\tikzset{snake it/.style={-stealth,
decoration={snake, 
    amplitude = .4mm,
    segment length = 2mm,
    post length=0.9mm},decorate}}
\usepackage{amssymb}

\usepackage[a4paper, total={6in, 10in}]{geometry}

\begin{document}
\title{\Large{\textbf{Covariant dynamics on the momentum space}}}
\date{}
\author{Boris Iveti\'c
\footnote{bivetic@yahoo.com}
  \\Wöhlergasse 6, 1100 Vienna, Austria \\}

\maketitle

\begin{abstract}
A geometrical interpretation of Schrödinger's kinetic and potential energy operators is proposed, allowing for a covariant momentum space formulation of the dynamics that is relevant for the theories with the deformation of the momentum space structure. Some specific examples are discussed in the context of flat space deformations and the Euclidean Snyder (spherical momentum space) model. In this formulation the dynamics for the deformations of the flat momentum space becomes trivial, while different versions of the Snyder model turn out to be dynamically equivalent. 
\end{abstract}

A number of quantum gravity scenarios predicts modification of continuum space(-time) picture at or near the Planck energy scale. Among these there is a class of models that arise from modification of the corresponding (energy-)momentum space. We shall be dealing in this letter with two such models. The first model arises from diffeomorphisms of the flat momentum space, leading to the deformation of Heisenberg algebra and a minimal uncertainty in the position \cite{vag1, vag2, vag3}. The second model, first introduced by Snyder \cite{sny}, postulates a spherical geometry of the momentum space.

Specifying a geometry/coordinatization of the momentum background does not uniquely determine the dynamics on this background.  In general approach one starts with a standard formulation of dynamics and seeks to generalize it, so as to incorporate a deformation of the space/momentum space geometrical structures, as well as the deformation of their symmetry (group) content. This can be done in more than one way. Additionally, it is a characteristic of Euclidean space that the points on the space can at the same time be treated as vectors, which does not hold on general spaces. All this is reflected in a large number of different proposals for the dynamics, a small selection of which is given in references \cite{gol, kady1,kady2, mir1, mir2, mign1, lus2, mign2}, which often give different results for what should correspond to same physical problems. 

Here we offer a novel approach to this problem. We start with the formulation of dynamics in the momentum representation in the standard case, and rewrite it in the form of geometrical invariants. This leads immediately to a natural formulation of the dynamics in both of the cases we consider, the deformation through the reparametrizations of the flat momentum space, as well as the deformation of the flat momentum space geometry itself, like in the Snyder model. To be specific, we deal only with a class of Schrödinger problems, but the guiding principles will also hold for more general theories, like quantum field theory (QFT), the details of which we postpone for a future work.
\\ \\
\textbf{1. Flat space diffeomorphisms} \\
An interesting and widely studied case that gives rise to the generalization of the uncertainty principle (GUP) arises from  redefinition of the momentum. This generalization starts with the standard momentum operator $\hat\eta_i$, where $i=1,2$ \footnote{We limit our analyses to two dimensions for simplicity, all our results and conclusions holding in the same way in three dimensions.}, and its adjoint position operator, $\hat x^i$, satisfying the canonical Heisenberg algebra,

\begin{equation}
[\hat x_i, \hat\eta^j]=i\hbar\delta^j_i, \ \ \ \ [\hat x_i, \hat x_j]=[\hat\eta^i, \hat\eta^j]=0,
\end{equation}
and from which the operators in the stationary Schrödinger equation are built,

\begin{equation}
(\hat K(\hat\eta_i)+\hat V(\hat x^i) ) \psi_n = E_n\psi_n,
\end{equation}
where $\hat K$ is the kinetic energy operator, and $\hat V$ is the potential energy operator. This is a spectral problem on an abstract Hilbert space. To be specific, one needs to choose the representation. If one chooses the momentum representation, assuming as usual Euclidean momentum space, with standard momentum coordinates, i.e. those in which the metric is Kroenecker's delta, then the momentum operator is just multiplication, and the position operator is gradient (a generator of infinitesimal translation on the Euclidean momentum space),

\begin{equation}
\hat\eta^i\to \eta^i, \ \ \ \hat x_i\to -i\hbar\frac{\partial}{\partial \eta^i}.
\end{equation}
Schrödinger's equation is then\footnote{Here and in rest of the paper a square means the sum of the squares of components, i.e. $\eta^2=\delta_{ij}\eta^i\eta^j$, and the repeated indices are summed.}

\begin{equation}\label{stand}
\left(\frac{\eta^2}{2m}+V\left(\frac{\partial}{\partial \eta^i}\right)\right) \psi_n (\eta) = E_n\psi_n(\eta),
\end{equation}

But what if one chooses a different set of coordinates, i.e. identifies the momentum with $p^i$, related to the original via

\begin{equation}\label{trans}
\eta^i=h(\alpha^2 p^2)p^i,
\end{equation}
where $\alpha$ is a constant of dimension inverse momentum and $h$ an arbitrary function? In this case, the momentum metric is

\begin{equation}
g^{ij}(p)=\frac{\partial \eta^i}{\partial p^k}\frac{\partial\eta^j}{\partial p^l}\delta^{kl}=h^2\delta^{ij}+4h'\alpha^2(h+\alpha^2p^2)p^i p^j
\end{equation}
and the position operator becomes

\begin{equation}\label{pos}
\hat x_i \to -i\hbar \frac{\partial p^j}{\partial\eta^i}\frac{\partial}{\partial p^j}=-i\hbar\left(\frac{1}{h}\frac{\partial}{\partial p^i}-p^i\frac{2\alpha^2h'}{h(h+2\alpha^2p^2h')}p^j\frac{\partial}{\partial p^j}\right),
\end{equation}
where $h'=\partial h/\partial(\alpha^2p^2)$, which are just the rules for tensor transformation ((2,0) tensor and a covector) upon the change of coordinates. 
%

 From (5) and (6) one obtains a deformed Heisenberg algebra
\begin{equation}
[\hat x_i, \hat p^j]=i\hbar\left(\frac{1}{h}\delta_i^j - \frac{2\alpha^2h'}{h(h+2\alpha^2p^2h')}\delta^i_kp^k p^j\right), \ \ \ \ [\hat x_i, \hat x_j]=[\hat p^i, \hat p^j]=0
\end{equation}
A special case of this transformation with $h=1/(1+\alpha^2p^2)$ was considered in \cite{vag1}, up to the leading order in $\alpha^2$, where it was shown to lead to the minimal uncertainty in the position $\Delta x\ge\hbar\alpha$.

The main question is how does the Schrödinger equation look like in this new setting? In \cite{vag1, vag2, vag3}, the kinetic energy operator was given as 

\begin{equation}
\hat K= \frac{p^2}{2m}
\end{equation}
But this choice is not unique. Different choices for the kinetic energy operator have been discussed in \cite{mign1} for the case of the Snyder model. In order to have a clear geometrical meaning of the kinetic energy operator, we define it as a geodesical distance from the origin divided with $2m$. In other words, the $\eta^2$ term in (\ref{stand}) is the square of the length of the shortest path from origin to the point $\eta$

\begin{equation}
\eta^2=\left( \int_0^\eta \sqrt{\delta_{ij} d\eta^i d\eta^j}      \right)^2=d^2(0,\eta),
\end{equation}
where $d(\eta, \eta')$ is the distance function (geodesic distance between points $\eta$ and $\eta'$), so that for a general variable $p$ defined through (\ref{trans}) the kinetic operator is

\begin{equation}\label{kin}
\hat K=\frac{1}{2m}\left( \int_0^p \sqrt{g_{ij} dp^i dp^j}      \right)^2=\frac{d^2(0,p)}{2m}= \frac{h^2p^2}{2m}
\end{equation}

 This definition is not new. In \cite{relloc1}, the definition of mass was given as the geodesic distance from the origin of the energy-momentum space, as a generalization of the usual $m^2=\eta_{\mu\nu}p^\mu p^\nu$, and the definition of kinetic energy was given as the square of the distance from a given point on the energy-momentum space to the point with the same mass but zero momentum, which reduces to our definition in the $c\to\infty$ limit.
The value of $p^2$ has a geometrical meaning only on the Euclidean space, so that the above proposal is the simplest and most natural one for the case of more general spaces, if one wishes to have a geometric interpretation of the kinetic energy operator.

What concerns the potential energy operator, we discuss its geometric form on two specific examples.
\\ \\
a) Example A: harmonic oscillator potential 
\\
In the standard case, the isotropic harmonic oscillator (HO) potential is 
\begin{equation}
\hat V_{HO}= \frac{m\omega^2}{2}\hat x^2= -\frac{m\omega^2\hbar^2}{2}\frac{\partial^2}{\partial \eta^2}
\end{equation}
We choose again to define it so as to give it a definite geometrical meaning. This is achieved by defining it as the divergence of the gradient, which for the general choice of coordinates on the flat space is

\begin{equation}\label{potho}
\hat V_{HO}= -\frac{m\omega^2\hbar^2}{2}\Delta=-\frac{m\omega^2\hbar^2}{2\sqrt {\text{det} g}}\frac{\partial}{\partial p^i} \left(\sqrt{\text{det} g} g^{ij}\frac{\partial}{\partial p^j}    \right)
\end{equation}
With the choices (\ref{kin}) and (\ref{potho}) for the kinetic and the potential energy operator the Schrödinger equation becomes completeley covariant, 
from whence it is clear  that, by construction, the eigenvalues remain the same independent of the choice of transformation $h$ in (\ref{trans}), so we may as well choose $h=1$. The only change between different coordinatizations is in the eigenvectors, which change according to

\begin{equation}\label{wftrans}
\psi_n(\eta)\to\psi_n(hp)
\end{equation}
which is just the rule for the transformation of the scalar function upon the change of coordinates.  The same procedure described here applies to any potential which is given in terms of powers of $\hat x^2$.
\\ \\
a) Example B: Coulomb's potential 
\\
An example of the potential that can not be expanded in terms of $\hat x^2$ is the Coulomb potential. It is an integral operator, which in the standard case looks like\footnote{We take $Ze^2/2\pi=1$.}

\begin{equation}
\hat V_{Coul}\psi(\eta)=\frac{1}{\hbar}\int\frac{\psi(\eta')d^2\eta'}{|\eta-\eta'|}. 
\end{equation}
To make this coordinate invariant (covariant), one generalizes in a natural way

\begin{equation}
d^2\eta\to d\Omega_p=\sqrt {\text{det} g}d^2p , \ \ \ \ \ \ |\eta - \eta'| \to d(p,p')=\sqrt{(h^2(\alpha^2p'^2)p'^2-h^2(\alpha^2p^2)p^2)^2},
\end{equation}
which in combination with (\ref{kin}) gives a fully covariant Schrödinger equation. The spectrum remains that of the usual hydrogen atom, with only wave functions changing according to (\ref{wftrans}).

A final note on the physical implications for the structure of space upon the transformation of the momentum space as in (\ref{trans}). It is argued that for some specific choice of transformation function, a "minimal length" emerges in the theory. We emphasize, however, that the position operator spectrum of a free particle remains continuous, as in the canonical case, a fact which does not dependend on the choice of the kinetic energy operator, but rather solely on the choice of the position operator (\ref{pos})

\begin{equation}
\hat x_i e^{i x_jhp^j/\hbar}=x_ie^{i x_jhp^j/\hbar}
\end{equation}
so no discretization of space emerges. The meaning of the uncertainty relation

\begin{equation}
\Delta x\ge \hbar\alpha,
\end{equation}
which follows from the $[\hat x, \hat p]$ for certain choices of the transformation $h$ \cite{vag3}, is,
from a strictly instrumentalist point of view, that upon simultaneous measurement of position and momentum, the position can be measured only up to a certain precision, regardles of the precision of the momentum measurement. In the same way, if in the canonical case one were to design an experiment to measure simultaneously position and a function $g(\alpha^2 p^2)p^i$ of the momentum, where $g$ is the inverse of the transformation function $h$, one would again arrive at the conclusion that there is a finite (non-vanishing) uncertainty in the measurment of the position in such experiment, irispective of the precision of the measurment of variable $gp^i$. This, however, would not imply the existence of a minimal length in the canonical setting. 
\\ \\ \\
\textbf{2. Euclidean Snyder space}\\
In this case we are considering an actual change of the geometry of the momentum background from flat to spherical. The momentum space is one (say upper) half of the 2-sphere, usually realised as embedded in the 3d Euclidean space

\begin{equation}\label{surf}
(\eta^1)^2+(\eta^2)^2+(\eta^3)^2=\beta^{-2}
\end{equation}
where $\beta$ is a constant of the dimension of inverse momentum, and with the origin at the north pole. There is again freedom in the choice of coordinatization of the sphere, or the choice of the physical momenta. Expressing the physical momentum coordinates $p^i$ in terms of the embedding space coordinates

\begin{equation}
\eta^i=h(\beta^2p^2)p^i, \ \ \ \ \ \ \eta^3=\sqrt{\beta^{-2}-h^2p^2},
\end{equation}
the momentum metric is given as

\begin{equation}
g^{ij}(p)=h^2\delta^{ij}+\beta^2\frac{4h'(h-\beta^2p^2h')-h^4}{1-\beta^2p^2h^2}p^ip^j
\end{equation}
and the position operator is 

\begin{equation}
\hat x_i=\beta\hat J_{i3}=-i\hbar\beta\frac{\sqrt{1-\beta^2p^2h^2}}{h}\left[ \frac{\partial}{\partial p^i} +\frac{2\beta^2h'}{h-2\beta^2p^2h'}\delta_{ik}p^k p^j\frac{\partial}{\partial p^j}  \right],
\end{equation}
which leads to a deformed Heisenberg algebra

\begin{equation}
[\hat x_i, \hat p^j]=i\hbar \frac{\sqrt{1-\beta^2p^2h^2}}{h}\left( \delta^j_i+ \frac{2\beta^2h'}{h-2\beta^2p^2h'}\delta_{ik}p^k p^j   \right), \ \ [\hat x_i,\hat x_j]=\beta^2\hat J_{ij}, \ \ [\hat p_i,\hat p_j]=0,
\end{equation}
where $\hat J_{12}$ is the angular momentum operator.
The benefit of our definitions is that all the geometrical considerations from the flat space case apply equally here. Kinetic energy operator is again given by

\begin{equation}\label{kins}
\hat K=\frac{d^2(0,p)}{2m}=\frac{1}{2m}\beta^{-2}\arccos^2 \sqrt{1-\beta^2p^2h^2}
\end{equation}
and we discuss the potential energy operator for the two cases:
\\ \\
a) Example A: harmonic oscillator potential 
\\
The divergence of the gradient (the Laplacian) is again given by (\ref{potho}).
which in combination with (\ref{kins}) gives the full Schrödinger equation,
for any choice of the coordinates. In a special case of one dimension, choosing $p=\mathcal P \theta$, with $\mathcal P\equiv\beta^{-1}$, as the momentum variable, for $\theta\in[-\pi/2, \pi/2]$ we have

\begin{equation}
\left(\frac{\mathcal P^2\theta^2}{2m}-\frac{m\omega^2\hbar^2}{2\mathcal P^2}\frac{\partial^2}{\partial\theta^2}\right)\psi_n(\theta)=E_n\psi_n(\theta)
\end{equation}
which upon the change of variable $k\equiv\mathcal P\theta/\sqrt{m\omega\hbar}$ becomes

 \begin{equation}\label{hos}
\left(k^2-\frac{\partial^2}{\partial k^2}\right) \psi_n(k)=\frac{2E_n}{\hbar\omega}\psi_n(k)
\end{equation}
with the boundary condition

 \begin{equation}
\psi_n\left(k=\pm \frac{\mathcal P\pi}{2\sqrt{m\omega\hbar}} \right)\to 0
\end{equation}

This has the same form as the 1d Schrödinger equation of the HO in the standard case, the only difference being in the boundary conditions, which also reduce to the standard ones in the limit $\mathcal P\to \infty$. This will be the feature of any 1d potential in the Snyder model, since it is characterized by the transformation of the phase space $(\eta, -i\hbar\partial/\partial \eta)\to(\mathcal P\theta, -i\hbar\partial/\partial (\mathcal P\theta))$, with the transformation of geometry manifesting itself only through the boundary conditions at finite points. This fact is not entirely surprising. In one dimension, the Heisenberg algebra reduces to 

\begin{equation}
[\hat x, \hat p]=i\hbar \frac{\sqrt{1-\beta^2p^2h^2}}{h}\left( 1+ \frac{2\beta^2h'}{h-2\beta^2p^2h'}p^2   \right), \ \ [\hat x,\hat x]= [\hat p,\hat p]=0
\end{equation}
without its characteristic noncommutativity between position operators and thus indistinguishable from the Heisenberg algebra appearing through the flat space reparametrization. This shows that one dimension does not suffice to capture  all of the features of the  Snyder model.

The eigenvalues of (\ref{hos}) are the same as in the standard case
\begin{equation}
E_n=\left(n+\frac{1}{2}\right)\hbar\omega
\end{equation}
with $n$ nonnegative integer, independent of the choice of momentum coordinates.
The eigenvectors are, up to normalization,  
\begin{equation}
\psi_n(k)=Re(D_n(\sqrt 2b)D_{-n-1}(i\sqrt 2 k)-D_n(\sqrt 2k)D_{-n-1}(i\sqrt 2 b)),
\end{equation}
for $n$ even and 
\begin{equation}
\psi_n(k)=Im(D_n(\sqrt 2b)D_{-n-1}(i\sqrt 2 k)-D_n(\sqrt 2k)D_{-n-1}(i\sqrt 2 b)),
\end{equation}
for $n$ odd, 
where $D_n(x)$ is the parabolic cylinder function and $b=\frac{\mathcal P\pi}{2\sqrt{m\omega\hbar}}$. 

This differs from the results for the 1d HO in the Snyder model obtained recently in \cite{mign2}, where different parametrizations were shown to lead to different eigenvalues. The discrepancy is traced back to the different choice for the kinetic energy operator taken in \cite{mign2}, where it was defined equal to $p^2/2m$ for any choice of coordinates, whereas in our model it is defined covariantly.

This is precisely the type of ambiguity in formulating the dynamics on general momentum backgrounds that was mentioned at the beginnig. Both of the choices of the kinetic energy operator, the one discussed here or the ones from e.g. \cite{vag3,mign2}, seem plausible and both reduce to the standard kinetic operator in the limit of vanishing deformation. Thus far nobody was able to derive the generalized dynamics from the first principles in such a convincing way that it would be accepted universally\footnote{Not the least for the lack of effort - see for instance the proposals to resolve this issue in \cite{relloc1, ac, carm}.}, probably because the first principles on which the formulation of dynamics in the standard case are grounded are unseparable from the Euclidean nature of the background on which it is usually defined. It might be that this question could be satisfactory resolved only upon the experimental reach. 
\\ \\
a) Example B: Coulomb potential 
\\
We apply the same recipe that we used in the case of flat momentum background, with 

\begin{align}
d^2\eta&\to d\Omega_p=\sqrt {\text{det} g}d^2p\\
 |\eta - \eta'|& \to d(p,p')=\beta^{-1}\arccos\left(\beta^2h_ph_{p'}p^ip'^i-\sqrt{1-\beta^2h_p^2p^2}\sqrt{1-\beta^2h_{p'}^2p'^2}    \right),
\end{align}
where $h_p=h(\beta^2p^2)$ and $h_{p'}=h(\beta^2p'^2)$, giving in combination with (\ref{kins}) a full Schrödinger equation in a covariant form.

A final note on the consequence of the generalization of the momentum space geometry  on the configurational space. In this case, unlike in the previous one,  the spectrum of the position operator is discrete. In one dimension, for instance, we have

\begin{equation}
\hat x e^{in\theta}=\hbar\beta n  e^{in\theta},
\end{equation}
where $n$ must be integer to enable the vanishing of the wave function at $\theta=\pm\pi/2$, which represent the points at infinity. This implies the emergence of the minimal length $\hbar\beta$ in the theory.  For a detailed exposition of the spatial lattice in three dimensions, we refer to \cite{lus1}.
\\ \\
\textbf{3. Outlook}\\
We have presented a geometrical interpretation of the kinetic and potential energy Schrödinger operators, allowing for a novel definition of the dynamics in case of nonstandard momentum space coordinates and/or geometry. The objection of this paper was to show the existence of covariant formulation of the dynamics, argue for its plausubility and deduce some of the consequnces that follow from it. This may be considered as a continuation of the program introduced in \cite{bor}, dubbed \textit{minimal extension principle}. Our principal findings are that in our construction, the dynamics from the flat space reparametrization is trivial, and that different realizations of the Snyder model are equivalent, which are both logical consequences of the covariance of our construction.

Considered deformations of the momentum space are expected to scale with the Planck energy\footnote{There is an emergent hope that some deformed models could be realisable at low energy, for instance analogly in Dirac materials, see e.g. \cite{alf}.}, calling for the definition of the dynamics in the context of QFT as well as special relativity. The basic principles laid down here for ordinary (non-relativistic) quantum mechanics will apply equally for the QFT, with momentum space being improved to energy-momentum space. The deformations of the flat energy momentum space will again lead to ordinary dynamics, while in case of Snyder the energy-momentum surface will be given by 
\begin{equation}
\left(\frac{E_\eta}{c}\right)^2-(\eta^1)^2-(\eta^2)^2-(\eta^3)^2=-\beta^{-2}
\end{equation}
which reduces to the momentum surface in (\ref{surf}) in the limit $c\to \infty$. All the generalizations will be along the lines presented, with the interchanges such as
\begin{equation}
\eta_{\mu\nu}p^\mu p^\nu\to d^2(0,p)
\end{equation}
and analogously. The details of this are to be given in a future work. 
\\
I am grateful to S. Mignemi for many discussions and for reading the manuscript.

\end{document}